\documentclass[aps,prd,twocolumn,groupedaddress]{revtex4-2}

\usepackage{amssymb}
\usepackage{amsmath}
\usepackage{graphicx}
\usepackage{enumerate}
\usepackage{mathtools}
\usepackage{color}
\usepackage{bbold}
\usepackage{subfigure}
\usepackage{slashed}
\usepackage{xcolor}
\usepackage{bm}
\usepackage{comment}

\begin{document}

\title{Geometric and Majorana phases in neutrino oscillations}

\author{Lucas Johns}
\email[]{NASA Einstein Fellow (ljohns@berkeley.edu)}
\affiliation{Departments of Astronomy and Physics, University of California, Berkeley, California 94720, USA}

\begin{abstract} 
Geometric (Aharonov--Anandan) phases in neutrino oscillations have been claimed [Phys. Lett. B 780 (2018) 216] to be sensitive to the Majorana phases in neutrino mixing. More recently, however, it has been pointed out [Phys. Lett. B 818 (2021) 136376] that the proposed phases are not gauge invariant. Using both kinematic and geometric approaches, we show that all gauge-invariant Aharonov--Anandan phases (including the off-diagonal geometric phases associated with flavor transitions) are independent of the Majorana phases. This finding, which generalizes the well-known fact that conventional oscillation experiments cannot discern the Dirac or Majorana nature of the neutrino, implies that a hypothetical interference experiment cannot distinguish between the two either.
\end{abstract}

\maketitle

\section{Introduction \label{sec:introduction}}

Measuring the parameters associated with neutrinos and their mixing is one of the major endeavors of particle physics. Many of these parameters are accessible through oscillation experiments, but not all. As is widely known, the nature of the neutrino---Dirac or Majorana---is reluctant to disclose itself. Strategies for making this discovery are sought, most prominently including neutrinoless double beta decay experiments \cite{dolinski2019}. Other ideas involve helicity--flavor oscillations \cite{gouvea2012, cirigliano2015, kartavtsev2015, dobrynina2016, chatelain2017, pustoshny2018, grigoriev2019} and neutrino decay \cite{balantekin2019, funcke2020}.

The question of interest in this paper is whether geometric phases in neutrino oscillations are sensitive to the Majorana phases even when the survival and transition probabilities are not. 

Connections between geometric phases and neutrino oscillations have been made from many different angles \cite{nakagawa1987, vidal1990, aneziris1991, smirnov1991, akhmedov1991, guzzo1992, naumov1992, naumov1994, blasone1999, wang2001, he2005, blasone2009, mehta2009, joshi2016, johns2017, joshi2017, dixit2018, joshi2020, capolupo2020, mukhopadhyay2020, manosh2021}, reflecting the diverse physics of geometric phases \cite{cohen2019}. Here the focus is on Aharonov--Anandan phases \cite{aharonov1987}, where the parametric variation is in the quantum state itself rather than the Hamiltonian. These phases appear naturally in neutrino oscillations and highlight the geometric significance of the mixing parameters.

Recent publications have disagreed on whether the Aharonov--Anandan phases in neutrino oscillations depend on the Majorana phases \cite{capolupo2018,lu2021,capolupo2021}. We consider this issue from two different perspectives (one kinematic, one geometric) and establish that none of the geometric phases are sensitive to the Majorana phases. This conclusion supports the contention of Ref.~\cite{lu2021} that the geometric phases of Ref.~\cite{capolupo2018}, which depend on the Majorana phases, are in fact gauge dependent and unphysical. 

In this paper it is made apparent, moreover, \textit{why} the geometric phases cannot be sensitive to the Majorana phases. In short, the freedom to rephase the charged-lepton fields renders the Majorana phases irrelevant to the geometry of projective Hilbert space. This is explicit in the usual parametrization of the mixing matrix $U$ [Eqs.~\eqref{eq:pmns2} and \eqref{eq:pmns} below], where the Majorana phases are equivalent to translations along the $U(1)$ fibers attached to the mass-eigenstate rays. Production and detection of neutrinos in flavor states ensure the unobservability of Majorana phases in traditional oscillation experiments \cite{giunti2010} \textit{and} in hypothetical neutrino interferometers.

Sec.~\ref{sec:phases} introduces geometric phases in the context of neutrino oscillations and describes the disagreement between Refs.~\cite{capolupo2018,capolupo2021} and Ref.~\cite{lu2021} in more detail. Sec.~\ref{sec:off} discusses what it means for geometric phases to be associated with flavor transitions. Sec.~\ref{sec:kinematic}, by expressing the geometric phase as the difference between the total and kinematic phases, then shows that the Majorana phases do not appear. Sec.~\ref{sec:geometric} confirms this finding by demonstrating that the Majorana phases are irrelevant to the geometry of projective Hilbert space. Sec.~\ref{sec:experiments} compares this situation with what one encounters in considering traditional oscillation experiments. Sec.~\ref{sec:discussion} concludes.

\section{Phases in neutrino oscillations \label{sec:phases}}

Neutrino oscillations are an interference phenomenon. A single particle's flavor state $| \nu \rangle$ is a superposition of energy eigenstates $| \nu_i \rangle$, each of which acquires phase at a different rate. These phases interfere with each other in a time-dependent way, causing flavor to oscillate.

Aside from the energy-eigenstate phases $e^{- i E_i t}$, there are phases and angles that are introduced by the conversion between mass and flavor bases,
\begin{equation}
| \nu (t) \rangle = \sum_i U^*_{\alpha i} e^{-i E_i t} | \nu_i \rangle,
\end{equation}
where the neutrino is assumed to have flavor $\alpha$ initially. With two flavors, the mixing matrix is
\begin{equation}
U = \begin{pmatrix} \cos\theta & \sin\theta \\ -\sin\theta & \cos\theta \end{pmatrix}\begin{pmatrix} e^{i\alpha_1} & 0 \\ 0 & 1 \end{pmatrix}, \label{eq:pmns2}
\end{equation}
with a single mixing angle $\theta$ and a single Majorana phase $\alpha_1$. With three flavors,
\begin{widetext}
\begin{equation}
U = \begin{pmatrix}
c_{12}c_{13} & s_{12}c_{13} & s_{13}e^{-i\delta_{CP}} \\
-s_{12}c_{23} -c_{12}s_{13}s_{23}e^{i\delta_{CP}} & c_{12}c_{23} - s_{12}s_{13}s_{23}e^{i\delta_{CP}} & c_{13}s_{23} \\
s_{12}s_{23} - c_{12}s_{13}c_{23}e^{i\delta_{CP}} & -c_{12}s_{23} -s_{12}s_{13}c_{23}e^{i\delta_{CP}} & c_{13}c_{23}
\end{pmatrix}
\begin{pmatrix}
e^{i\alpha_1} & 0 & 0 \\
0 & e^{i\alpha_2} & 0 \\
0 & 0 & 1
\end{pmatrix}
\label{eq:pmns}
\end{equation}
\end{widetext}
with mixing angles $\theta_{12}$, $\theta_{13}$, and $\theta_{23}$, the Dirac phase $\delta$, and the Majorana phases $\alpha_1$ and $\alpha_2$.

Thus we have two sources of phases: the dynamics and the basis transformation. Neutrino flavor transition probabilities depend on the dynamical parameters ($\delta m_{ij}^2 / 2E$) as well as some of the parameters in $U$ ($\theta_{ij}$ and $\delta$). The probabilities are notoriously independent of $\alpha_1$ and $\alpha_2$.

But quantum mechanics exhibits other kinds of phases, namely geometric ones, which are only observable under special experimental conditions. How do these relate to the other phases in neutrino oscillations? Is it possible that \textit{they} depend on the Majorana phases, making neutrino oscillations potentially sensitive to $\alpha_1$ and $\alpha_2$ even if the standard experiments are not?

Capolupo \textit{et al.} \cite{capolupo2018} have argued that this is indeed the case. They consider the Aharonov--Anandan phase \cite{aharonov1987, mukunda1993}
\begin{equation}
\Phi^\Gamma = \textrm{arg} \left\langle \psi (\mathbf{s}_1) | \psi (\mathbf{s}_2) \right\rangle - \textrm{Im} \int_{\mathbf{s}_1}^{\mathbf{s}_2} d\mathbf{s} \left\langle \psi (\mathbf{s}) \bigg| \frac{d}{d\mathbf{s}} \bigg| \psi (\mathbf{s}) \right\rangle
\end{equation}
along the path $\Gamma$ from $\mathbf{s}_1$ to $\mathbf{s}_2$, and first establish that
\begin{equation}
\Phi^\Gamma_{\nu_\alpha} (z) = \textrm{arg} \left\langle \nu_\alpha (0) | \nu_\alpha (z) \right\rangle - \textrm{Im} \int_{0}^{z} dz' \left\langle \nu_\alpha (z') | \dot{\nu}_\alpha (z') \right\rangle \label{eq:diagkin}
\end{equation}
is independent of the two-flavor Majorana phase $\alpha$ for $\alpha = e, \mu$. (The dot indicates derivative with respect to propagation distance $z'$.) They then observe that
\begin{equation}
\Phi^\Gamma_{\nu_\alpha \rightarrow \nu_\beta} = \textrm{arg} \left\langle \nu_\alpha (0) | \nu_\beta (z) \right\rangle - \textrm{Im} \int_{0}^{z} dz' \left\langle \nu_\alpha (z') | \dot{\nu}_\beta (z') \right\rangle
\end{equation}
does depend on $\alpha_1$ for $\beta \neq \alpha$. The authors refer to these quantities as the geometric phases associated with flavor transitions.

The claims of Ref.~\cite{capolupo2018} have recently been disputed by Lu \cite{lu2021}, who observes that $\Phi^\Gamma_{\nu_\alpha \rightarrow \nu_\beta}$ is not gauge invariant for $\beta \neq \alpha$ and that $\alpha_1$ can be eliminated from the formula by a charged-lepton field rotation. Capolupo \textit{et al.} \cite{capolupo2021} have replied, asserting that Lu's arguments are incorrect and maintaining that $\Phi^\Gamma_{\nu_\alpha \rightarrow \nu_\beta}$ is gauge invariant and measurable.

Ref.~\cite{capolupo2021} claims to prove the gauge invariance of $\Phi^\Gamma_{\nu_e \rightarrow \nu_\mu}$ by showing, in that paper's Eq.~11, that the phase is invariant under $| \nu_{e,\mu} \rangle \longrightarrow e^{i \lambda} | \nu_{e,\mu} \rangle$. This is inadequate as a proof because $| \nu_e \rangle$ and $| \nu_\mu \rangle$ can be rotated independently (see the next section). It can immediately be seen that $\Phi^\Gamma_{\nu_e \rightarrow \nu_\mu}$ is \textit{not} invariant under $| \nu_{e,\mu} \rangle \longrightarrow e^{i \lambda_{e,\mu}} | \nu_{e,\mu} \rangle$ when $\lambda_e \neq \lambda_\mu$.

A crucial point of disagreement between Ref.~\cite{lu2021} and Refs.~\cite{capolupo2018,capolupo2021} is whether charged-lepton field rephasing can legitimately be used to transform the mixing matrix. The argument made by the latter authors is that the mixing matrices
\begin{gather}
U^{(1)} = \begin{pmatrix} \cos\theta & \sin\theta e^{i \alpha_1} \\ -\sin\theta & \cos\theta e^{i \alpha_1} \end{pmatrix}, \notag \\
U^{(2)} = \begin{pmatrix} \cos\theta & \sin\theta e^{i \alpha_1} \\ -\sin\theta e^{- i \alpha_1} & \cos\theta \end{pmatrix} \label{eq:defu1u2}
\end{gather}
identify physically distinct scenarios despite being related by
\begin{equation}
U^{(2)} = \begin{pmatrix} 1 & 0 \\ 0 & e^{-i \alpha_1} \end{pmatrix} U^{(1)}. \label{eq:u1u2}
\end{equation}
The authors claim that charged-lepton field rephasing cannot be used to transform these matrices because the mixing matrix is properly defined by
\begin{equation}
\left( U^{(j)} \right)^\dagger H^{(j)} U^{(j)} = \begin{pmatrix} m_1 & 0 \\ 0 & m_2 \end{pmatrix} \label{eq:hambases}
\end{equation}
for $j = 1,2$. By this line of reasoning, $U^{(1)}$ is the appropriate matrix for Dirac neutrinos, which have
\begin{equation}
H^{(1)} = \begin{pmatrix} m_{ee} & m_{e\mu} \\ m_{e\mu} & m_{\mu\mu} \end{pmatrix},
\end{equation}
and $U^{(2)}$ is the appropriate matrix for Majorana neutrinos, which have
\begin{equation}
H^{(2)} = \begin{pmatrix} m_{ee} & m_{e\mu} e^{i \alpha_1} \\ m_{e\mu} e^{- i \alpha_1} & m_{\mu\mu} \end{pmatrix}.
\end{equation}

The problem with this argument is that the mixing matrix is \textit{not} defined by Eq.~\eqref{eq:hambases}. It is defined (for two flavors) by
\begin{align}
&\mathcal{L}_{CC} (x) = \notag \\
&~~~~~\frac{g}{\sqrt{2}} \sum_{\alpha = e, \mu} \sum_{i = 1,2} \overline{\alpha_L} (x) \gamma^\rho U_{\alpha i} \nu_{i L} (x) W_\rho (x) + \textrm{H.c.}, \label{eq:cc}
\end{align}
where $g$ is the weak coupling constant and $e_L$, $\mu_L$, $\nu_L$, and $W_\rho$ are respectively the electron, muon, neutrino, and $W$ boson fields. The matrices $U^{(1)}$ and $U^{(2)}$ correspond to different bases, not different physics. If one wishes to work in the flavor basis, then $U^{(j)}$ is equal to the mixing matrix $U$ defined by Eq.~\eqref{eq:cc}---up to whatever rephasing freedom $U$ has---because this is precisely what one means by \textit{flavor basis}. Since $U$ can be transformed by rephasing $e_L$ and $\mu_L$, $U^{(j)}$ has the associated freedom as well. That is, contrary to the assertions of Refs.~\cite{capolupo2018, capolupo2021}, $U^{(1)}$ and $U^{(2)}$ are physically equivalent if they are related by multiplication by a diagonal matrix of phases. As we see from Eq.~\eqref{eq:u1u2}, that is the case here. 

The remarks above support the conclusion of Ref.~\cite{lu2021} that $\Phi^\Gamma_{\nu_\alpha \rightarrow \nu_\beta}$ is not a gauge-invariant geometric phase. But this statement on its own does not prove that geometric phases are \textit{necessarily} insensitive to Majorana phases: perhaps other quantities can be constructed that exhibit the desired dependence.

\section{Off-diagonal geometric phases \label{sec:off}}

Although not mentioned in Refs.~\cite{capolupo2018,lu2021,capolupo2021}, gauge-invariant quantities do exist that are associated with transitions. They are known as off-diagonal geometric phases \cite{manini2000, mukunda2001, hasegawa2001, filipp2003, wong2005}. The set of all geometric phases---diagonal and off-diagonal---gives all of the geometric information about neutrino oscillations. Thus, to see if there is sensitivity to the Majorana phases, it suffices to evaluate the off-diagonal geometric phases. We do this in the next section.

But before doing so, let us consider why the diagonal phases on their own do not exhaust the geometric information. Suppose that a beam of two-flavor neutrinos is split in two and recombined in such a way that at the end of the process the flavor state is
\begin{equation}
| \nu (z) \rangle = \left[ a_0 (z) | \nu_e \rangle + b_0 (z) | \nu_\mu \rangle \right] + \left[ a(z) | \nu_e \rangle + b(z) | \nu_\mu \rangle \right],
\end{equation}
where the coefficients are properly normalized and depend on the paths (and the potentials encountered along them). In this hypothetical neutrino interferometry experiment, the second set of square brackets encloses the part of the beam that is rerouted over a different path than the one taken by the rest of the incident beam. Further suppose that $| \nu (0) \rangle = | \nu_e \rangle$. Then the probability of measuring electron flavor at position $z$ is
\begin{equation}
P_{\nu_e \rightarrow \nu_e} (z) = | \langle \nu_e | \nu (z) \rangle |^2 = | a_0 (z) + a(z) |^2.
\end{equation}
The survival probability is a measure of one of the \textit{diagonal} phases because the overlap is being calculated with respect to the initial flavor state. The expression $| a_0 (z) + a(z) |^2$ is sensitive to the phase difference acquired by the neutrinos along the two paths, but only in the $\nu_e$ part. Because of the orthogonality of the flavor states, some of the phase information is not discoverable by this procedure.

That information can be discovered through the transition probability
\begin{equation}
P_{\nu_e \rightarrow \nu_\mu} (z) = | \langle \nu_\mu | \nu (z) \rangle |^2 = | b_0 (z) + b(z) |^2, \label{eq:transprob}
\end{equation}
which is sensitive to the relative phase acquired by a neutrino as it oscillates into $\nu_\mu$ along the two paths. Here the overlap is taken with respect to a state that is orthogonal to the initial one, hence the phase that arises is \textit{off-diagonal}.

In the literature on off-diagonal geometric phases, one typically works in terms of energy eigenstates and imagines that the Hamiltonian is changed such that eigenstate $| \psi_i \rangle$ becomes an orthogonal eigenstate $| \psi_j \rangle$. Then, in the course of this evolution, the diagonal geometric phase becomes undefined because one needs the argument of $\langle \psi_i | \psi_j \rangle = 0$. Phase information cannot be lost, however, because $| \psi_j \rangle$ could very well be subsequently brought back to $| \psi_i \rangle$. In a similar way, if $|\nu (0) \rangle = | \nu_e \rangle$ evolves into $| \nu (z) \rangle = |\nu_\mu \rangle$, then $\langle \nu (0) | \nu(z) \rangle = \langle \nu_e | \nu_\mu \rangle = 0$ entails phase ambiguity. If the state only partially evolves into $\nu_\mu$, some of the phase information is nonetheless still missing if one only considers the inner product with $| \nu_e \rangle$.

But while the diagonal phase $\langle \nu_e | \nu (z) \rangle$ is gauge invariant, the off-diagonal quantity $\langle \nu_\mu | \nu(z) \rangle$ is not. One way to understand why they differ in this respect is as follows. Consider the gauge transformation
\begin{equation}
| \nu (0) \rangle \longrightarrow e^{i \varphi} | \nu (0) \rangle.
\end{equation}
Then
\begin{equation}
| \nu_e \rangle \longrightarrow e^{i \varphi} | \nu_e \rangle
\end{equation}
and the diagonal phase is invariant under the change:
\begin{equation}
\langle \nu_e | \nu (0) \rangle \longrightarrow \langle \nu_e | \nu (0) \rangle.
\end{equation}
But a rotation of $| \nu_\mu \rangle$ is not entailed because there is no $| \nu_\mu \rangle$ part of the initial flavor state. Thus
\begin{equation}
\langle \nu_\mu | \nu (0) \rangle \longrightarrow e^{i \varphi} \langle \nu_\mu | \nu (0) \rangle.
\end{equation}
In this set-up, the relative phase between the production state $| \nu_e \rangle$ and the measurement state $| \nu_\mu \rangle$ is unphysical. Each of these states has its own $U(1)$ gauge invariance. In the literature on off-diagonal geometric phases, a similar statement is that
\begin{equation}
| \psi_{i,j} \rangle \longrightarrow e^{i \varphi_{i,j}} | \psi_{i,j} \rangle,
\end{equation}
which is to say that each eigenstate can be rephased independently.

Gauge invariance motivates constructing off-diagonal geometric phases as \cite{manini2000}
\begin{equation}
\gamma_{ij}^\Gamma = \sigma_{ij}^\Gamma \sigma_{ji}^\Gamma
\end{equation}
with
\begin{equation}
\sigma_{ij}^\Gamma = \Phi \left[ \langle \psi_i^{\|} (\mathbf{s}_1) | \psi_j^{\|} (\mathbf{s}_2) \rangle \right]
\end{equation}
and $\Phi [z] = z / |z|$ for $z \neq 0$. The states here are the parallel-transported vectors
\begin{equation}
| \psi_i^{\|} ( \mathbf{s}_2 ) \rangle = \exp \left[ - \int_{\Gamma} d\mathbf{s} \cdot \left\langle \psi_i (\mathbf{s}) | \nabla_\mathbf{s} \psi_i (\mathbf{s}) \right\rangle \right] | \psi_i (\mathbf{s}_2) \rangle,
\end{equation}
The exponential factor, which is required to ensure parallel transport along $\Gamma$, cancels the dynamical phase
\begin{equation}
i \varphi_{i,\textrm{dyn}}^\Gamma = \int_{\Gamma} d\mathbf{s} \cdot \left\langle \psi_i (\mathbf{s}) | \nabla_\mathbf{s} \psi_i (\mathbf{s}) \right\rangle
\end{equation}
accumulated during the evolution. Therefore
\begin{equation}
\sigma_{ij}^\Gamma = \exp \bigg[ i ~\textrm{arg} \langle \psi_i (\mathbf{s}_1) | \psi_j (\mathbf{s}_2) \rangle - i \varphi_{j,\textrm{dyn}}^\Gamma \bigg]
\end{equation}
and the off-diagonal geometric phase is
\begin{align}
\gamma_{ij}^\Gamma = \exp \bigg[ i ~\textrm{arg} &\left( \langle \psi_i (\mathbf{s}_1) | \psi_j (\mathbf{s}_2) \rangle \langle \psi_j (\mathbf{s}_1) | \psi_i ( \mathbf{s}_2) \rangle \right) \notag \\
- &i \varphi_{i,\textrm{dyn}}^\Gamma - i \varphi_{j,\textrm{dyn}}^\Gamma \bigg]. \label{eq:gammakin}
\end{align}
Considering the same gauge transformations as before,
\begin{equation}
\sigma_{ij}^\Gamma \longrightarrow e^{i (\varphi_j - \varphi_i )} \sigma_{ij}^\Gamma
\end{equation}
and $\gamma_{ij}^\Gamma$ is clearly gauge invariant.

Eq.~\eqref{eq:gammakin} is a kinematic expression, but the off-diagonal phase can also be understood in a more obviously geometric manner as the integral of the Berry curvature 2-form. In Sec.~\ref{sec:geometric} we adopt a geometric perspective to show that none of the Aharonov--Anandan phases can depend on the Majorana phases. The important point for now is that the surface in projective Hilbert space is bordered by curves connecting four rays: those associated with $| \psi_i (\mathbf{s}_1) \rangle$, $| \psi_j (\mathbf{s}_1) \rangle$, $| \psi_i (\mathbf{s}_2) \rangle$, and $| \psi_j (\mathbf{s}_2) \rangle$. In a similar way, $P_{\nu_e \rightarrow \nu_\mu} (z)$ [Eq.~\eqref{eq:transprob}] involves $| \nu (0) \rangle = | \nu_e \rangle$, $|\nu_\mu \rangle$, and the two states (corresponding to the two paths) that are superposed at position $z$ to give $| \nu (z) \rangle$. 

Having established the nature of the transition phases, we show in the next two sections that neither these nor the diagonal geometric phases are sensitive to the Majorana phases.

\section{The kinematic approach \label{sec:kinematic}}

The kinematic approach identifies the geometric phase as the total accumulated phase minus the dynamical part. It is straightforward to show that neither the total nor the dynamical phase can depend on the Majorana phases and that, therefore, the geometric phase cannot either.

Considering first the initial--final overlap, we have
\begin{align}
\langle \nu_\alpha (0) | \nu_\beta (z) \rangle &= \left( \sum_i U_{\alpha i} \langle \nu_i | \right) \left( \sum_j U_{\beta j}^* | \nu_j \rangle e^{-i E_j z} \right) \notag \\
&= \sum_i U_{\alpha i} U_{\beta i}^* e^{-i E_i z}. \label{eq:overlap}
\end{align}
Since $U_{\alpha i}$ contains a factor $e^{i \alpha_i}$ and $U_{\beta i}^*$ contains a factor $e^{-i \alpha_i}$, the Majorana phases drop out for all $\alpha, \beta$. ($\alpha_2 = 0$ in the two-flavor case and $\alpha_3 = 0$ in the three-flavor case.) 

Next we consider the dynamical phases. Here we have, using $| \nu (0) \rangle = | \nu_\alpha \rangle$,
\begin{align}
i \varphi_\textrm{dyn} (z) &= \int_0^z dz' \langle \nu (z') | \dot{\nu} (z') \rangle \notag \\
&= - i \left( \sum_i | U_{\alpha i} |^2 E_i \right) z. \label{eq:phidyn}
\end{align}
Again the Majorana phases clearly drop out.

Since the total and dynamical phases are independent of the Majorana phases, the geometric phases (diagonal and off-diagonal) must be as well, as per Eq.~\eqref{eq:diagkin} for the diagonal phases and Eq.~\eqref{eq:gammakin} for the off-diagonal phases.

\section{The geometric approach \label{sec:geometric}}

With two flavors, the flavor state is a vector $| \nu \rangle$, which can be parametrized as
\begin{equation}
| \nu \rangle = e^{i \chi} \begin{pmatrix} \cos \frac{\vartheta}{2} e^{i \frac{\phi}{2}} \\ \sin \frac{\vartheta}{2} e^{-i \frac{\phi}{2}} \end{pmatrix}. \label{eq:2psiparam}
\end{equation}
Here $\vartheta$ and $\phi$ are polar angles, parametrizing the position of the associated ray in projective Hilbert space (which for two-level systems is the Bloch sphere). The phase factor $e^{i \chi}$ parametrizes the $U(1)$ fiber. While in this sense it has a geometric meaning, it is irrelevant in projective Hilbert space. Indeed, the associated ray is the density matrix $\rho = | \psi \rangle \langle \psi |$, from which $\chi$ obviously cancels.

Similarly, the Euler-angle parametrization of SU(3) is convenient when there are three flavors \cite{byrd1999}:
\begin{equation}
| \nu \rangle = e^{i \chi} \begin{pmatrix} \sin\vartheta \cos\beta e^{i (\eta + \gamma)} \\ \sin\vartheta \sin\beta e^{-i(\eta - \gamma)} \\ \cos\vartheta \end{pmatrix}. \label{eq:3psiparam}
\end{equation}
Rays in the three-level projective Hilbert space are parametrized by four angles. One parameter again simply parametrizes the fiber.

We work in the mass basis because we would like to consider the evolution of $| \nu \rangle$. For three flavors, we have
\begin{equation}
| \nu_\alpha \rangle = \tilde{U}^*_{\alpha 1} e^{-i \alpha_1} | \nu_1 \rangle + \tilde{U}^*_{\alpha 2} e^{-i \alpha_2} | \nu_2 \rangle + \tilde{U}^*_{\alpha 3} | \nu_3 \rangle,
\end{equation}
where $\tilde{U}$ is the PMNS matrix with Majorana phases set to zero. We then parametrize each $| \nu_i \rangle$ with its own set of angles $\lbrace \chi_i, \eta_i, \gamma_i, \beta_i, \vartheta_i \rbrace$. Then
\begin{widetext}
\begin{align}
| &\nu_\alpha \rangle = \notag \\
&\tilde{U}^*_{\alpha 1} e^{i ( \chi_1 - \alpha_1)} \begin{pmatrix} \sin \vartheta_1 \cos \beta_1 e^{i (\eta_1 + \gamma_1)} \\ \sin\vartheta_1 \sin\beta_1 e^{-i(\eta_1 - \gamma_1)} \\ \cos\vartheta_1 \end{pmatrix} + \tilde{U}^*_{\alpha 2} e^{i ( \chi_2 - \alpha_2)} \begin{pmatrix} \sin \vartheta_2 \cos \beta_2 e^{i (\eta_2 + \gamma_2)} \\ \sin\vartheta_2 \sin\beta_2 e^{-i(\eta_2 - \gamma_2)} \\ \cos\vartheta_2 \end{pmatrix} + \tilde{U}^*_{\alpha 3} e^{i \chi_3} \begin{pmatrix} \sin \vartheta_3 \cos \beta_3 e^{i (\eta_3 + \gamma_3)} \\ \sin\vartheta_3 \sin\beta_3 e^{-i(\eta_3 - \gamma_3)} \\ \cos\vartheta_3 \end{pmatrix}. \label{eq:parametrized}
\end{align}
\end{widetext}
Each of $\chi_1, \chi_2, \chi_3$ can be chosen freely, so let $\chi_1$ and $\chi_2$ absorb $\alpha_1$ and $\alpha_2$, respectively. Then the Majorana phases have in effect been absorbed into the mass eigenstates. It follows that the Majorana phases are, like the $\chi$ parameters, geometrically irrelevant in projective Hilbert space: they cannot show up in geometric phases. An analogous argument applies with two flavors.

Note that if an \textit{arbitrary} initial state could be produced, then the Majorana phases could potentially be detectable. For example, if one could make a measurement of
\begin{align}
&\left| \langle \nu_e | \left( A_1 | \nu_1 \rangle + A_2 | \nu_2 \rangle + A_3 | \nu_3 \rangle\right) \right|^2 = \notag \\
&~~~~~~~~~~~~~~~\left| A_1 \tilde{U}_{e 1} e^{i \alpha_1} + A_2 \tilde{U}_{e 2} e^{i \alpha_2} + A_3 \tilde{U}_{e 3} \right|^2,
\end{align}
where $A_i$ is an arbitrary coefficient, then such an experiment would be sensitive to $\alpha_1$ and $\alpha_2$. The dependence, however, would be unrelated to geometric phases, or even to the dynamics at all. The reason the Majorana phases are not \textit{geometric} is that the PMNS matrix can be parametrized in such a way that the phases attach to mass eigenstates. That is, $U$ can be factorized into the product of a rotation matrix (possibly including a Dirac phase) and a diagonal matrix of Majorana phases. The reason they are not \textit{observable} in oscillation experiments is that neutrinos are always produced and detected in flavor states. 

The argument above is essentially that the Majorana phases can be absorbed into the mass eigenstates, precluding the phases from showing up in the dynamics. We come to the same conclusion by considering the quantum geometric tensor \cite{provost1980, berry1989, braunstein1994}:
\begin{equation}
T_{\mu\nu} = \langle \partial_\mu \psi | \partial_\nu \psi \rangle - \langle \partial_\mu \psi | \psi \rangle \langle \psi | \partial_\nu \psi \rangle,
\end{equation}
where $\mu, \nu$ refer to coordinates in parameter space. This object is invariant under the usual gauge transformations. Its real and imaginary parts are both geometrically significant. Writing
\begin{equation}
T_{\mu\nu} = g_{\mu\nu} + i \frac{V_{\mu\nu}}{2},
\end{equation}
the real part $g_{\mu\nu}$ is the metric tensor measuring distances between rays in projective Hilbert space and the imaginary part $V_{\mu\nu}$ is the curvature 2-form whose flux through a circuit $C$ accounts for the geometric phase $\gamma (C)$. From the parametrizations of $\psi$ in Eqs.~\eqref{eq:2psiparam} and \eqref{eq:3psiparam}, it is clear that neither $g_{\mu\nu}$ nor $V_{\mu\nu}$ has any dependence on the $\chi$ phases. 

The implication, as before, is not that Majorana phases cannot show up at all. The implication is that if they do, it is because of the production and detection processes, not the geometry. Here we have another perspective on the geometric irrelevance of Majorana phases, this time from the vantage point of the quantum geometric tensor rather than that of the states.

The arguments in this section are also consistent with a simple counting of degrees of freedom. A state has $U(1)^n$ gauge invariance for $n$ flavors because each $| \nu_i \rangle$ can be rephased independently due to orthogonality. [This is what allows for the $\chi$ parameters to be chosen freely in Eq.~\eqref{eq:parametrized}.] Mixing is described by an $n \times n$ unitary matrix, and $\textrm{dim}[U(n) / U(1)^n] = n(n-1)$. Given production and detection in flavor states, charged-lepton field rephasing is equivalent to the further freedom to choose the phases of $| \nu_\mu \rangle$ and $| \nu_\tau \rangle$ relative to $| \nu_e \rangle$. This leaves a total of $(n-1)^2$ parameters, none of which is a Majorana phase.

\section{Interference vs. oscillation experiments \label{sec:experiments}}

We are now in a position to say that the Majorana phases are inaccessible to interference experiments for the same reasons that they are inaccessible to standard oscillation experiments.

In an oscillation experiment, the relevant measurement is
\begin{equation}
P_{\nu \rightarrow \nu_e} (z) = | \langle \nu_e | \nu (z) \rangle |^2,
\end{equation}
where, following Giunti's treatment in Ref.~\cite{giunti2010}, we let
\begin{equation}
| \nu (0) \rangle = A_e | \nu_e \rangle + A_\mu | \nu_\mu \rangle + A_\tau | \nu_\tau \rangle
\end{equation}
with arbitrary coefficients $A_\alpha$. Then
\begin{equation}
| \nu (z) \rangle = \sum_i \left( A_e U^*_{e i} + A_\mu U^*_{\mu i} + A_\tau U^*_{\tau i} \right) | \nu_i \rangle
\end{equation}
and
\begin{align}
P&_{\nu \rightarrow \nu_e} (z) \notag \\
&= \left| \sum_{i,j} \langle \nu_j | U_{ej} \left( A_e U^*_{e i} + A_\mu U^*_{\mu i} + A_\tau U^*_{\tau i} \right) e^{-i E_i z} | \nu_i \rangle \right|^2 \notag \\
&= \left| \sum_i \left( A_e | U_{ei} |^2 + A_\mu U^*_{\mu i} U_{ei} + A_\tau U^*_{\tau i} U_{ei} \right) e^{- i E_i z} \right|^2.
\end{align}
Because $\alpha_1$ and $\alpha_2$ appear only in phase factors multiplying their respective mass states, they cancel out just as they did in Eqs.~\eqref{eq:overlap} and \eqref{eq:phidyn}. This occurs at the amplitude level.

It is possible to be led to a seemingly different result by choosing the mixing matrix differently \cite{giunti2010}. Going back to two flavors, consider the change $U^{(1)} \longrightarrow U^{(2)}$, where these are the matrices defined in Eq.~\eqref{eq:defu1u2}. As discussed in Sec.~\ref{sec:phases}, $U^{(1)}$ and $U^{(2)}$ are related by rephasing of the charged-lepton fields and must lead to physically equivalent results. But using $U^{(2)}$, one calculates that $\alpha_1$ does not vanish from $P_{\nu \rightarrow \nu_e}$ because
\begin{equation}
\left( U^{(2)}_{\mu 1} \right)^* U^{(2)}_{e 1} = - \left( U^{(2)}_{\mu 2} \right)^* U^{(2)}_{e 2} = - \sin\theta \cos\theta e^{i \alpha_1}.
\end{equation}
(In this case $A_\tau = 0$ by assumption of having only two flavors.)

The resolution is that the $\alpha_1$-dependent terms vanish when only one of the coefficients $A_\mu$ is nonzero. This condition is necessitated by having production in flavor states. (Ref.~\cite{giunti2010} also addresses situations involving neutral-current processes.) We saw in the previous section that the same point applies to geometric phases. Ultimately it is production and detection in flavor states that account for the inaccessibility of Majorana phases by interference \textit{or} oscillation experiments.

\section{Discussion \label{sec:discussion}}

The findings reported above are unchanged if neutrino oscillations take place in a medium where the mass states do not coincide with the energy eigenstates. Geometric phases of Berry type \cite{berry1984}, where the Hamiltonian is parametrically varied (by changing the matter density, for example), offer no advantage in sensitivity to the Majorana phases as compared to geometric phases of Aharonov--Anandan type.

Majorana phases \textit{can} appear, however, due to helicity--flavor oscillations, which take place in a larger Hilbert space. Neutrinos produced at weak-interaction vertices are not definite-helicity states. As is well known, though, the associated effects are suppressed by $m_\nu / E_\nu$, the ratio of neutrino mass to neutrino energy. This small factor is what makes it so challenging to determine whether neutrinos are Majorana or Dirac. Unfortunately, geometric phases do not present a strategy for discerning the nature of the neutrino without incurring the usual $m_\nu / E_\nu$ penalty.

\begin{acknowledgements}
This work was supported by NASA through the NASA Hubble Fellowship Grant Number HST-HF2-51461.001-A awarded by the Space Telescope Science Institute, which is operated by the Association of Universities for Research in Astronomy, Incorporated, under NASA contract NAS5-26555.
\end{acknowledgements}

\bibliography{all_papers}

\end{document}